# Multiferroic Decorated Fe$_2$O$_3$ Monolayer Predicted from First Principles

Jing Shang [1], Chun Li [2,3], Xiao Tang [1], Aijun Du [1], Ting Liao [1], Yuantong Gu [1], Yandong Ma [4], Liangzhi Kou [1]*, Changfeng Chen [5]*

[1] *School of Mechanical, Medical and Process Engineering, Queensland University of Technology, Brisbane, QLD 4001, Australia*

[2] *School of Mechanics, Civil Engineering and Architecture, Northwestern Polytechnical University, Xi'an 710072, China*

[3] *Department of Mechanical Engineering, University of Manitoba, Winnipeg MB R3T 5V6, Canada*

[4] *School of Physics, State Key Laboratory of Crystal Materials, Shandong University, Jinan 250100, China*

[5] *Department of Physics and Astronomy, University of Nevada, Las Vegas, Nevada 89154, United States*

**Corresponding Author**

*E-mail: liangzhi.kou@qut.edu.au

*E-mail: chen@physics.unlv.edu

**ABSTRACT.** Two-dimensional (2D) multiferroics exhibit cross-control capacity between magnetic and electric responses in reduced spatial domain, making them well suited for next-generation nanoscale devices; however, progress has been slow in developing materials with required characteristic properties. Here we identify by first-principles calculations robust 2D multiferroic behaviors in decorated $Fe_2O_3$ monolayer, showcasing Li@$Fe_2O_3$ as a prototypical case, where ferroelectricity and ferromagnetism stem from the same origin, namely Fe $d$-orbit splitting induced by the Jahn-Teller distortion and associated crystal field changes. These findings establish strong materials phenomena and elucidate underlying physics mechanism in a family of truly 2D multiferroics that are highly promising for advanced device applications.

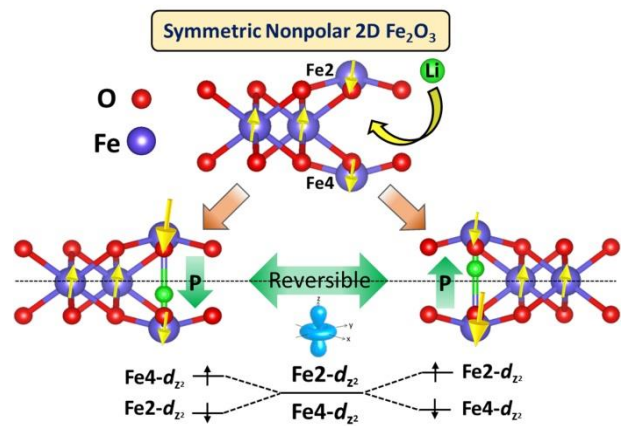

**Introduction**

Ferromagnetism (FM) and ferroelectricity (FE) are fundamental physics phenomena underlying designs and working principles in many advanced nanoscale devices. Multiferroic materials that exhibit simultaneous presence and effective cross control of magnetic and electric polarization hold great promise for many applications. Of particular interest are atomically thin two-dimensional (2D) multiferroic materials [1-7] that are ideally suited for device miniaturization. Depending on the coupling between FE and FM orders, multiferroics can be classified into two types. For type I, FE polarization generally originates from either spontaneous atomic displacements [8-10] or synergistic effects of charge and orbital orders [5, 11], while FM stems from partially filled $d$ shells of transition metals. These materials usually possess weak magnetoelectric coupling and ineffective cross control between FM and FE, thus limiting their functional capacity. For type-II multiferroics, FE polarization is derived from local magnetic order [1] [12] [13] with much stronger magnetoelectric coupling. One example is $Hf_2VC_2F_2$ monolayer that exhibits FE due to broken inversion symmetry originating from an antiferromagnetic (AFM) order [1]. Another example is $VOX_2$ (X = Cl, Br, and I) monolayer [6] that possesses both FM and FE derived from the $d_{xy}$ orbital splitting of the V cation. There have been hitherto only a few type-II multiferroics discovered, and more are urgently needed for diverse material platforms.

Iron oxides, such as $Fe_3O_4$ [14] and $Fe_2O_3$ [15], show multiple magnetic states, and recent works have revealed the presence of multiferroicity in Fe-O systems [16-18] 2D $Fe_3O_4$ and $\varepsilon$-$Fe_2O_3$ magnetic films display ferroelectric switching [18-19], but the origin of FE polarization is unclear while the magnetoelectric coupling has not been fully explored. Very recently, α-$Fe_2O_3$ monolayer has been exfoliated from its bulk matrix and demonstrated to be a magnetic semiconductor [15], but this material is non-ferroelectric due to its centrosymmetric structure. Multiferroicity has be induced in α-$Fe_2O_3$ monolayer by building α-$Fe_2O_3$/$BaTiO_3$ ferromagnetic/ferroelectric heterostructures, and magnetism in 2D α-$Fe_2O_3$ is tunable by the interaction with $BaTiO_3$ [16], but the magnetoelectric coupling is weak because the FM and FE order are driven by different sources and mechanisms.

In this study, we present results from first-principles studies that show robust multiferroicity in Li-decorated α-$Fe_2O_3$ (Li@α-$Fe_2O_3$) monolayer. The symmetry breaking by the decorating Li atoms between the iron atoms induces a Jahn-Teller distortion that produces FE and FM order, both stemming from the $d_{z^2}$-orbit splitting due to changes in the crystal field. Similar behaviors also have been found in α-$Fe_2O_3$ monolayer decorated by oxygen atoms, introducing a new family of 2D multiferroic materials. These findings provide insights into fundamental physics processes and mechanisms underlying the coexisting FE and FM order in atomically thin multiferroics and expand the material offerings for practical implementation.

**Computational method**: We have performed calculations using the Vienna Ab-initio Simulation Package (VASP) [20]. The PAW approach is adopted to describe the ion-electron interaction [21]. The electron exchange-correlation energy was treated by the PBE functional [22], and the energy cutoff for the planewave expansion is set to 600 eV. The calculations with the PBE+U functional have been conducted to check the effective on-site Coulomb interaction of $3d$ electrons of Fe, which plays a significant role in determining the electronic properties. To better describe van der Waals forces, zero damping DFT-D3 (where D stands for dispersion) method with the Grimme vdW correction [23] is adopted. By minimizing the stresses and forces, the lattice vectors and atomic positions are fully relaxed until the energy differences are converged to within $10^{-6}$ eV with a force convergence threshold of $10^{-3}$ eV/Å for all the calculations except for the calculation about electric field. The Brillouin zone is represented by the Monkhost-Pack special k-point mesh [24] of $17 \times 17 \times 1$. The band structures of $Fe_2O_3$ monolayers are calculated along the special lines of M → G → K → M. We use the Gaussian smearing with 0.1 meV width in the Brillouin zone integration. The energy of polarization transformation for Li-decorated $Fe_2O_3$ monolayers is calculated with climbing-image nudged elastic band (CI-NEB) method [25]. The vacuum space is set to at least 20 Å to avoid the interaction between periodic images.

**Results and discussion.** The $Fe_2O_3$ monolayer is modelled based on the structure exfoliated from iron ore hematite (α-$Fe_2O_3$), which exhibits ferromagnetic order [15]. We first establish benchmarks for pure $Fe_2O_3$ monolayer, whose unit cell [Figure 1(a)] contains 4 Fe atoms with 4 possible magnetic configurations depending on the spin orientations, FM, $AFM_1$, $AFM_2$, and $AFM_3$ [see Figure S1 (a-d)]. GGA+U calculations with $U_{eff}$=2.0 ~ 5.0 eV indicate that $AFM_2$ is always the ground state regardless of the $U_{eff}$ values (see Figure S1 (e) and Table S1 for details), where Fe2 and Fe4 have the same spin direction, opposite to that on Fe1 and Fe3 [see Figure 1(a)]. Due to different bonding environments (Fe1 is bonded with 6 surrounding O atoms while Fe2 is bonded with 3 O atoms), the spin polarization on these Fe atoms are different (4.17 μB on Fe1, -3.98 μB on Fe2), leading to an overall magnetic moment of 0.213 μB per unit cell and a ferrimagnetic ground state. This structure is a semiconductor with a band gap of 0.937 eV, see Figure S1 (f), which is consistent with recent experiments [15]. The ferrimagnetic ground state also has been confirmed by recent theoretical and experimental works [26]. Pure $Fe_2O_3$ monolayer possesses the $C_{2h}$ structural symmetry, which renders same electron distributions on Fe atoms at the upper and lower surfaces (Fe2 and Fe4), and the whole system thus has zero out-of-plane electric polarization. This symmetry is also reflected by the $e_g$ orbit of Fe2 and Fe4 with the equal band splitting, see Figure 3(c) and bottom panel (schematic diagram) of Figure 1(a). Therefore, pure $Fe_2O_3$ monolayer is a non-polar ferrimagnetic semiconductor.

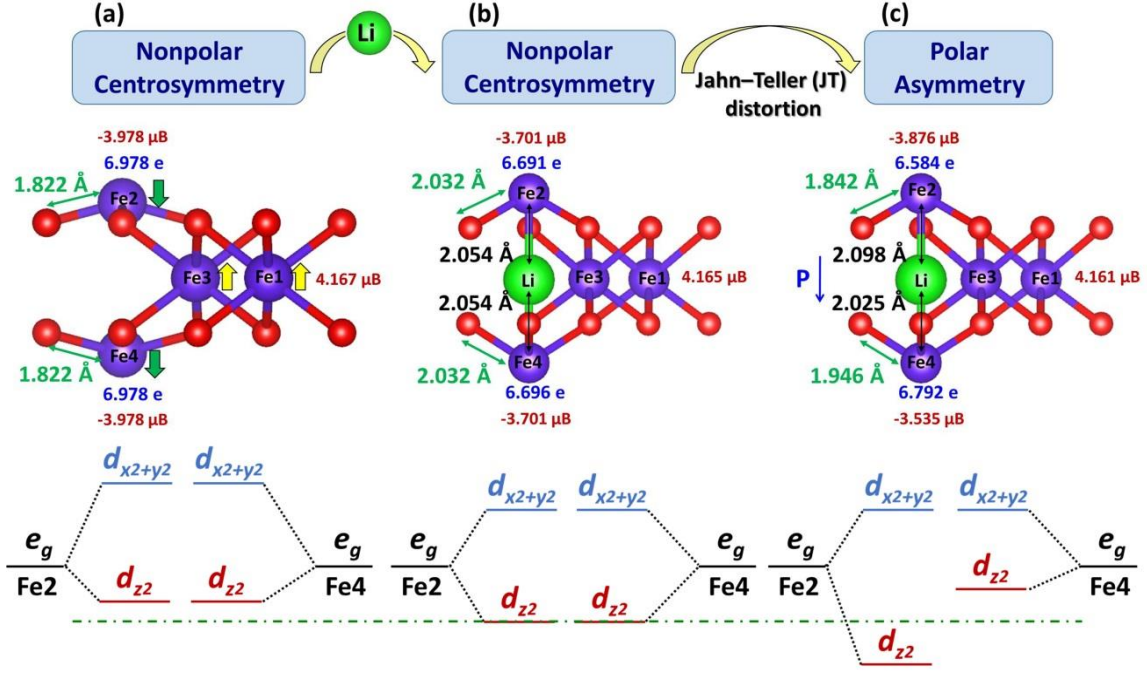

Figure 1. The monolayer structures and spin configurations (yellow and green arrows) of (a) centrosymmetric pure $Fe_2O_3$, (b) centrosymmetric $Li@Fe_2O_3$ monolayer, and (c) the asymmetric structure when Jahn-Teller (JT) distortion occurs. Indicated are bond lengths, charge distributions and magnetic moments (μB) on Fe2 and Fe4 atoms at $U_{eff}$ = 4.0 eV.

Exfoliated $Fe_2O_3$ monolayer possesses a large number of dangling bonds, and passivation of these bonds may induce spontaneous polarization in the out-of-plane direction, possibly generating ferroelectricity that coexists with the intrinsic ferrimagnetism. Below we focus on ferroelectricity induced by Li decoration and will examine this as a more general phenomenon toward the end of the paper. We first assess structural stability of Li-decorated $Fe_2O_3$ monolayer by calculating the formation energy,

$$E_f = (E_{Li@Fe_2O_3} - E_{pure-Fe_2O_3} - E_{Li})/n$$

where $E_{Li@Fe_2O_3}$, $E_{pure-Fe_2O_3}$ and $E_{Li}$ are the per unit-cell total energies of $Li@Fe_2O_3$ monolayer, pure $Fe_2O_3$ monolayer, and a single isolated lithium atom, respectively, and $n$ is the number of atoms in the system. The calculated $E_f$ is -0.188 eV/atom, indicating an exothermic reaction when Li atom is embedded into the $Fe_2O_3$ monolayer. This result shows that fabrication of $Li@Fe_2O_3$ monolayer is energetically feasible. To further examine the structural stability, we also checked the $E_f$ by using $E_{Li}$ as energy of bulk Li, the formation energy is still negative, -0.04 eV/atom.

The fully relaxed structure with a Li atom embedded between Fe2 and Fe4 is shown in Figure 1(c), and the original symmetry is obviously broken. To understand the structural and electronic evolutions, we first built a symmetric structure by placing a Li atom in the middle of the Fe2-Fe4 position to preserve the inversion symmetry, as shown in Figure 1(b). The uniform initial Fe-O and Fe-Li bond lengths ensure a centrosymmetric structural configuration with equivalent charge and magnetic moment distributions on Fe2 and Fe4 atoms, resulting in zero net out-of-plane polarization. As shown in Figure 1(a) and (b), Li-atom incorporation stretches the Fe-O bond length from 1.822 Å to 2.032 Å, leading to redistributions of electron and spin polarization. However, this symmetric structure is only metastable. When the initially imposed symmetry constraint is removed, structural relaxation produces a deformed monolayer with an energy that is 0.2 eV/unit-cell lower than the symmetric structure. For the distorted Li@$Fe_2O_3$, the bond lengths of the upper and lower surfaces Fe2-O both become shorter (1.842 and 1.946 Å) but unequal. The Li atom is obviously shifted along the z direction, and the Fe-Li bond lengths are 2.098 and 2.025 Å, respectively, compared with 2.054 Å for the symmetric case, as shown in Figure 1(b-c). The Jahn-Teller (JT) distortion [27] occurring spontaneously in this system helps to lower the total energy and break structural symmetry. As a result, charge distributions on Fe2 and Fe4 atoms (6.584 e and 6.792 e, respectively) from Bader analysis are unequal, resulting in the misalignment of the centres of positive and negative charges. The electric polarization is estimated to be -0.188 e·Å from the berry phase calculation at U=4.0 eV. To check the effect of the Li atom on the magnetic ground state, we calculated the total energy of all the magnetic arrangements (FM, $AFM_1$, $AFM_2$, and $AFM_3$) similar to pure $Fe_2O_3$ monolayer. The results show that the $AFM_2$ spin configuration remains the ground state (see Figure 2) regardless of the U value. However, the distribution of spin charges is affected by the structural symmetry breaking and becomes unequal as seen in Figure 1(c). Similar to the origin of out-of-plane polarization, the magnetism is also resulted from the $e_g$ orbit splitting of Fe atoms. The coexisting ferroelectricity and ferrimagnetism makes Li-doped $Fe_2O_3$ monolayer a true 2D multiferroic.

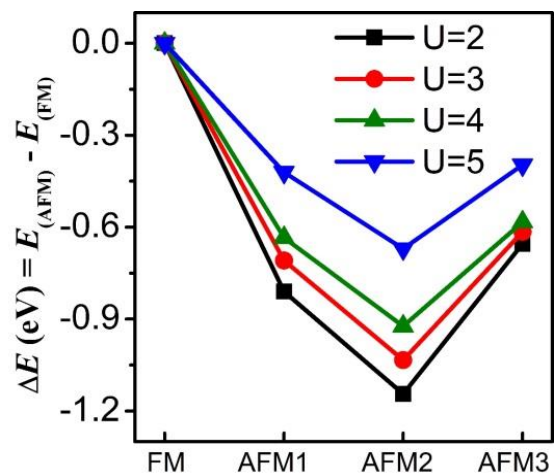

Figure 2. The calculated energy differences (ΔE) between AFM and FM spin orders with various $U_{eff}$ (Fe) from 2.0 to 5.0 eV. The total energies of FM are set to 0. The magnetic ground state is confirmed to be $AFM_2$ spin order at all studied $U_{eff}$ (Fe) values.

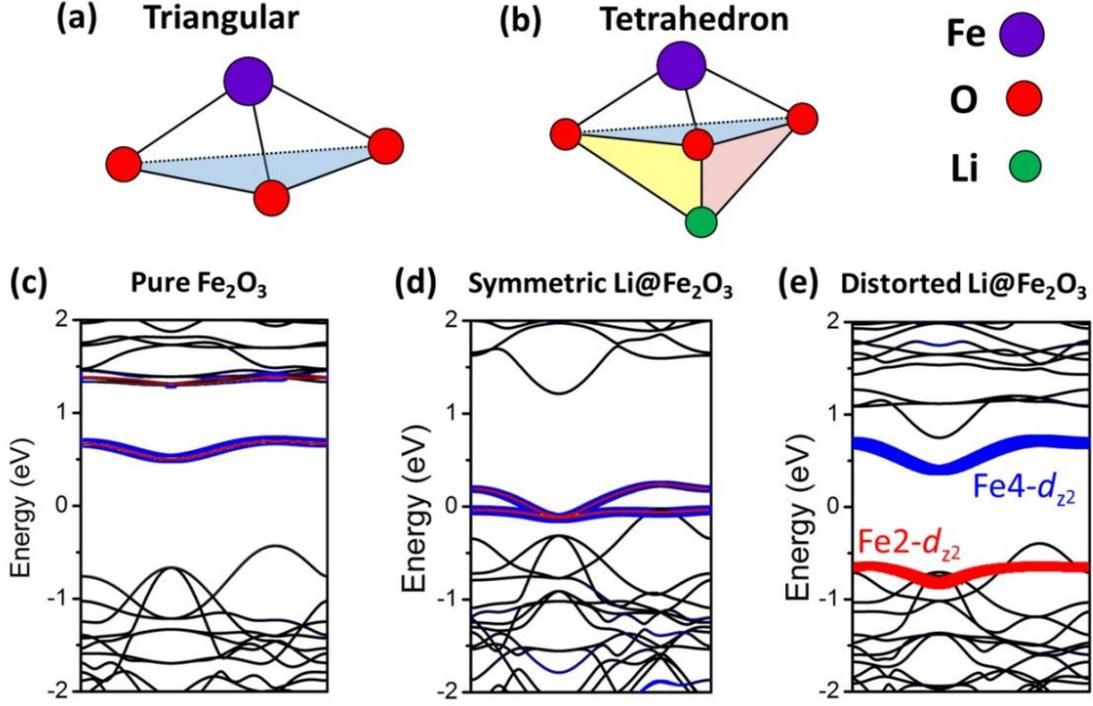

Figure 3. (a, b) In-plane triangular and tetrahedral ligand fields. The projected band for Fe2 and Fe4 atoms in (c) pure $Fe_2O_3$ monolayer, (d) symmetric Li@$Fe_2O_3$ monolayer, and (e) asymmetric Li@$Fe_2O_3$ monolayer with JT distortion. The red and blue lines in (c-e) indicate the $d_{z^2}$ orbitals projection of Fe2 and Fe4 atoms.

To explore the mechanism underlying the distortion-induced out-of-plane FE polarization of Li@$Fe_2O_3$ monolayer, we examine changes of the crystal field and energy reduction from the associated electronic band splitting. Here, we again compare the symmetric $Fe_2O_3$, symmetric Li@$Fe_2O_3$, and asymmetric Li@$Fe_2O_3$ monolayers and assess the Fe-$d_{z^2}$ orbital splitting in the distinct Fe-O and Fe-O-Li crystal-field environments in Figure 3(a-b). For pure $Fe_2O_3$ monolayer, Fe atoms at the surface (Fe2 and Fe4) are coordinated with three surrounding oxygen atoms, forming the in-plane triangular ligand field, and the Fe-$e_g$ states of both Fe2 and Fe4 split equally into the higher $d_{x^2+y^2}$ and lower $d_{z^2}$ states stemming from the symmetric structures [see Figure 1(a)]. For the symmetric Li@$Fe_2O_3$ case, the crystal field changes from in-plane triangle to tetrahedron, and consequently $e_g$ orbitals of both Fe2 and Fe4 atoms are reduced significantly to lower energy levels, as indicated in Figure 3(c-d). Once the symmetry constrain is removed, a spontaneous electronic energy split between the Fe2- and Fe4-$d_{z^2}$ states [red and blue lines in Figure 3(e)] with orbital-ordering occurs, which originates from the JT distortion with the elongated and shortened Fe-Li bond lengths, thereby inducing the higher Fe4-$d_{z^2}$ and lower Fe2-$d_{z^2}$ electronic energy levels. The asymmetric shift of the $d$ orbitals, especially the $d_{z^2}$ states of Fe2 and Fe4 atoms not only leads to the out-of-plane electric polarization, but also induces spin redistributions on relevant Fe atoms, as shown in Figure 1(c), to reach the ferrimagnetic ground state. Here, both the electric polarization and magnetic states share the same origin, stemming from the *d*-orbital shift after Li decoration.

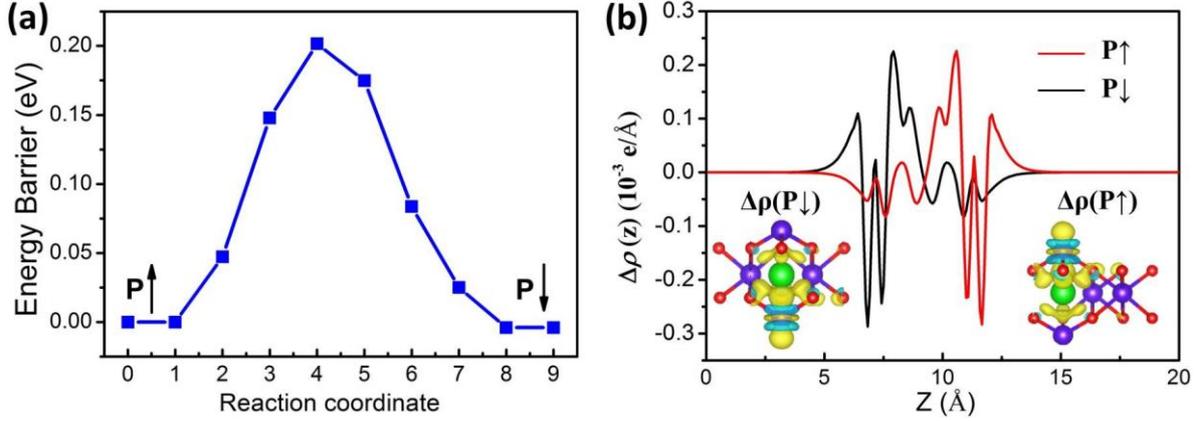

Figure 4. The activation energy barrier (a) and the differential charge density (b) for FE polarization up (red) and down (black) of the distorted Li@Fe$_2$O$_3$ monolayer way at $U_{eff}$ = 4.0 eV. The isosurface value in (b) is set to 0.005 e Å$^3$.

The energy barrier for the polarization reversal gives good indications on FE stability and switchability. We have performed CI-NEB calculations to assess the energy barrier of polarization reversal at selected U values for Li@Fe$_2$O$_3$ monolayer. Results in Figure 4 show that the transition barrier is 0.202 eV/unit cell at $U_{eff}$ = 4.0 eV, indicating that the ferroelectric state is stable and the switch can be manipulated under external stimuli. Correspondingly, the out-of-plane polarization continuously changes from -0.18 e·Å to +0.18 e·Å from berry phase calculations, which indicates that there is no depolarization during the FE reversal. In contrast to the double-well barrier seen in ferroelectric CuMP$_2$X$_6$ (M = Cr, V; X = S, Se) [28], a single-well barrier is observed in Li@Fe$_2$O$_3$ as shown in Figure 4 (a), and this result is attributed to the unstable structure of paraelectric phase and spontaneous structural distortion that would occur when the constrain of centrosymmetric Li@Fe$_2$O$_3$ monolayer is removed. To provide a solid evidence of the polarization reverse, we plotted the differential charge density as shown in Figure 4(b). It indicates the polarization in Li@Fe$_2$O$_3$ monolayer is reversed where the internal polarization towards up and down is switched with obvious charge gain and lose around Fe2 atom, respectively.

It is noted that multiferroic behaviors are not limited to the case discussed above, but a more general phenomenon in decorated Fe$_2$O$_3$ monolayer. In fact, the JT distortion and induced FE polarization is also observed when O atom is inserted between the iron atoms. Similar multiferroic behaviours in Fe$_2$O$_3$ monolayer is observed with introduction of an O atom between Fe2 and Fe4 although the JT distortion and associated FE polarization are relatively weak, see Figure S2.

**Conclusion.** In summary, we have shown by first-principles calculations that Li@Fe$_2$O$_3$ monolayer is a robust 2D multiferroic with concurrent out-of-plane FM and FE orders. The electric polarization in Li@Fe$_2$O$_3$ monolayer stems from the crystal-field splitting of $d_{z^2}$ orbitals in Fe2 and Fe4 atoms, driven by a spontaneous Jahn-Teller distortion induced by the decorated Li atoms. Meanwhile, a net out-of-plane magnetization on Fe2 to Fe4 is induced by the FE polarization, highlighting the common origin and close correlation of the FM and FE responses to external stimuli. The multiferroic behaviour unveiled in the present work is a general phenomenon in multiple decorated Fe$_2$O$_3$ monolayer systems, providing new material platforms for exploring intriguing physics mechanisms and innovative device applications.

**Author information**

Corresponding Author

*E-mail: liangzhi.kou@qut.edu.au

*E-mail: chen@physics.unlv.edu

**Author Contributions**

The manuscript was written through contributions of all authors. All authors have given approval to the final version of the manuscript.

**Notes**

The authors declare no competing financial interest.

**Acknowledgement**

We acknowledge the grants of high-performance computer time from computing facility at the Queensland University of Technology, the Pawsey Supercomputing Centre and Australian National Computational Infrastructure (NCI). L.K. gratefully acknowledges financial support by the ARC Discovery Project (DP190101607).

C.L. acknowledges the financial supports from the National NSF (Grant No. 11872309) of China and the Fundamental Research Funds for the Central Universities (Grant No. 3102017JC01003) of China.